\documentclass[a4paper,11pt]{article}
\pdfoutput=1 % if your are submitting a pdflatex (i.e. if you have
\usepackage{jcappub} % for details on the use of the package, please
\usepackage[T1]{fontenc} % if needed
\usepackage{float}
\usepackage[utf8]{inputenc}
\usepackage{hyperref}
\usepackage{cleveref}

\title{\boldmath Possible Nature of Dark Matter}
\author[a,1]{Wasif Husain\note{Corresponding author.}}
\author[a]{Anthony W. Thomas}
\affiliation[a]{ARC Centre of Excellence for Dark Matter Particle Physics and CSSM, Department of Physics, University of Adelaide, SA 5005, Australia}
\emailAdd{wasif.husain@adelaide.edu.au}
\emailAdd{anthony.thomas@adelaide.edu.au}
\abstract{We present a study of neutron star models that contain dark matter (DM) in the core. The DM is assumed to have a particle nature and to be self-interacting. Using constraints on the mass and radius of neutron stars, we investigate the allowed properties of either bosonic or fermionic DM particles. We consider cases where it constitutes up to 15\% of the mass of the star, even though conventional mechanisms cannot generate such large fractions. For this purpose three different models of neutron stars are considered, the first involving nucleons only, the second including hyperons, and the last involving strange matter in the core. Different EoSs are constructed for the various cases of fermionic and bosonic DM. These EoSs are solved for selected properties of the DM particles and the results are tested against mass, radius and tidal deformability constraints for neutron stars. The distribution of energy density of DM and ordinary matter inside the neutron stars is also presented. It is found that if the DM is fermionic in nature it does not just sit in the core but it is present everywhere in the star, from the centre to outside the surface and may even envelop it.}

\begin{document}
	\maketitle
	\flushbottom
	
	\section{Introduction}
	\label{sec:intro}
	Dark matter and its non interacting nature with Standard Model particles has been a mystery since decades. Its presence can be felt only because of gravity on a astrophysical scale and the key to solve the dark matter mystery lies in measuring the non-gravitational interactions of dark matter with ordinary particles. Ordinary matter contributes approximately 5\% of the total energy content of the universe while the dark matter contributes approximately 1/4 of the total content of the universe and remaining the 70\% is dark energy. The structure of galaxies and their evolution depend strongly on the nature of dark matter. The understanding of dark matter helps understand the role of particle physics in the evolution of the universe. 
	\par
	Neutron stars may be viewed as astrophysical laboratories to test and enhance our knowledge of behaviour of nuclear matter at the extremes of pressure, energy density, temperature. They have also been used to test proposals concerning the nature of DM \cite{Nelson:2018xtr,Ellis_2018,Kouvaris_2015,Mukhopadhyay_2017,Li_2012}. In this study the nature of DM is explored and they are used to put a constraint on the mass and self-interaction of dark matter particles for each type of dark matter, fermionic or bosonic. It is assumed that the dark matter has a particle nature and self-interacting. DM self-interactions have been used in many different studies to address various astrophysical problems \cite{Rocha_2013,Zavala_2013,Peter_2013,Spergel_2000,Faraggi_2002,Wandelt:2000ad,Mohapatra_2002,Kusenko_2001,Loeb_2011,Kouvaris_2012,2013MNRAS.430.1722V,Kaplinghat_2014,Petraki_2014,Schutz_2015}.For example, a DM-DM self-scattering cross-section per DM mass in the 0.1 - 10 cm$^2$/g range, helps solve the cusp vs core problem and the too big to fail" problem. There are some constraints on DM self-interactions that put a limit on the strength of DM  self-interactions. DM self-interactions should not be strong enough to destroy the elliptic shape of spiral galaxies \cite{Feng_2009,Feng_2010} or remove the sub-clusters of the bullet cluster \cite{Burgess_2001}.
	\par The abundance of dark matter in the universe may effect the properties of the compact astrophysical objects like white dwarfs, neutron stars, black holes etc, as they capture dark matter particles \cite{1978AJ.....83.1050S}.
	Some other models used with compact astronomical objects to study DM are given in Refs. \cite{PhysRevD.77.043515,PhysRevD.77.023006,PhysRevD.40.3221,f2b55bdf3fc1466081f4c0d091544a2d,PhysRevLett.105.011301,Lopes_2012,Taoso_2010,CIARCELLUTI201119,Sandin_2009,Leung:2011zz,B_hmer_2007}. Some of these models assume dark matter as ideal gas of self-interacting Fermions and suggest a non-self-annihilating condensate of dark matter \cite{Li_2012,L_2012} present at the core of compact astronomical objects. In addition, Ref. \cite{PhysRevD.81.123521} suggests that non-annihilating DM particles having a mass greater than 1 GeV contained at the core of neutron star will collapse the neutron stars. References \cite{Mukhopadhyay_2017, Kouvaris_2012, PhysRevD.82.063531, Li_2012} have used similar values (1 GeV) for the mass of DM particles and selected a DM self-interaction that helps solving the cusp vs core and the too big to fail problem. Therefore in this article we have also assumed that the DM particles has a mass 1 GeV while the fermionic dark photon (fermionic DM exchange particle) has a mass 100 MeV \cite{Mukhopadhyay_2017} and bosonic DM self-interaction scattering length is 1 fm \cite{Li_2012}.
	\par The central idea of this research is that the neutron stars accrete DM, and that will settle in the core. After settling the dark matter in the core  it would form a degenerate dark matter core inside the neutron star and that may collapse the neutron star. These captured dark matter particles contribute to the total mass of the neutron stars and will change their properties. Based on these effects we may put a constraint on the dark matter particles. If dark matter is bosonic in nature then according to Bose-Einstein condensation theory it must condense in the core below a critical temperature. How much dark matter a neutron star can accommodate depends on various factors, including the nature of dark matter, such as whether the dark matter is bosonic or fermionic and whether it is strongly or weakly self interacting. Other recent studies of the effects may be found in Refs. \cite{Nelson:2018xtr,Mukhopadhyay_2017,Kouvaris_2015,Ellis_2018}.
	\par In section 2, the EoSs employed to model the neutron stars are explained. In this section EoSs for both types of DMs are defined along with the EoSs for nuclear matter. In section 3, the structural equations are calculated by modifying the TOV equations to make them suitable for application to two fluids. In section 4, the EoSs and structural equations are integrated simultaneously and results are shown. This is followed by the conclusion section, where the results are discussed and conclusions regarding the nature of DM.  
	\section{Equation of state}
	To examine the effects of dark matter on neutron stars, three examples of neutron star core structures are considered, neutron stars made of only nuclear matter, hyperons and strange matter. For pure nuclear matter and hyperons at the core we have considered N-QMC700 (pure nuclear matter EoS) and F-QMC700 (EoS includes hyperons) which we generated using the  Quark-Meson coupling model
	\cite{PhysRevLett.93.132502}, as given in Ref \cite{RIKOVSKASTONE2007341} while for strange matter we have considered the EoS based on the MIT bag model \cite{PhysRevD.10.2599}, and given in Ref \cite{Urbanec_2013}. In addition, we need an EoS for self-interacting asymmetric fermionic dark matter and self-interacting bosonic dark matter. We have tested both dark matter EoSs with each of the neutron star EoSs with the combination solved using the modified TOV equation \cite{Tolman169,PhysRev.55.374} for two fluids. Finally as shown in \cite{Husain_2020}, the properties of neutron stars are sensitive to the EoS in the lower energy density region, typically at the crust. There we have used the Baym-Bethe-Pethick (BBP) EoS \cite{BAYM1971225} for the energy density region below 100 MeV/fm$^3$.
	\subsection{Self-interacting Asymmetric Fermionic Dark Matter (AFDM)} For self-interacting asymmetric dark matter the approach given in Refs. \cite{Kouvaris_2015,Mukhopadhyay_2017} is adopted. This uses a similar approach to vector meson exchange in the hadronic matter with the result ($\hbar = 1, c = 1$)
	\begin{equation}
	\epsilon_\chi = {m_\chi^4} \chi(x) + \frac{1}{(3\pi^2)^2}\frac{x^6m_\chi^6}{m_I^2},
	\end{equation}
	\begin{equation}
	P_\chi = {m_\chi^4} \phi(x) + \frac{1}{(3\pi^2)^2}\frac{x^6m_\chi^6}{m_I^2},
	\end{equation}
	where $m_I$ is the mass of the self interacting particle,  $\chi$(x) and $\phi$(x) are functions of the Fermi momentum, given by $x = \frac{(3\pi n_\chi)^{1/3}}{m_\chi}$, with $n_\chi$ the dark matter number density. 
	\begin{equation}
	\chi(x) = \frac{1}{8\pi^2}(x\sqrt{1+x^2}(1+2x^2)- \log(x+\sqrt{1+x^2})),    
	\end{equation}
	\begin{equation}
	\phi(x) = \frac{1}{8\pi^2}(x\sqrt{1+x^2}(2x^2/3 - 1)+ \log(x+\sqrt{1+x^2})) ,
	\end{equation}
	Equations (2.1) and (2.2) together make the EoS. The first term in both equations comes from the fermionic nature (follows the Pauli exclusion principle), while the send term represents the self interaction nature of the AFDM.  If asymmetric fermionic dark matter is not self interacting then it will be a gas of fermionic dark matter particles following the Pauli exclusion principle, unlike the bosonic dark matter and the second term from the equations (2.1) and (2.2) will vanish. As suggested in \cite{Kouvaris_2012}  asymmetric fermionic DM with attractive Yukawa-type  self-interactions can form destructive black holes at the core of old neutron stars which puts a constraint on the AFDM EoS.
	\subsection{Self-interacting Bosonic Dark Matter}
	In this case we suppose that the dark matter is a self-interacting dilute particle gas. Since we have considered DM is bosonic in nature, it must follows the Bose-Einstein statistics and below a certain temperature  all bosonic dark matter particles must condense to a same quantum state. At this temperature a coherent state of particles evolve. The critical temperature in terms of energy density is given by $T_{cr}$ = $\frac{2\pi l_a}{m_\chi^{5/3}k_B}\epsilon_\chi^{2/3}$, where m$_\chi$ is the mass of the dark particle in the condensate, $k_B$ is Boltzmann’s constant, and $\epsilon_\chi$ is the energy density. For a dilute gas of bosonic particles at such low temperature, only binary collisions at low energy are relevant. These are characterized by a scattering length ($l_\chi$) parameter. Details can be found in Refs.\cite{Li_2012,L_2012} . Precisely the EoS is given by
	\begin{equation}
	P_\chi = \frac{2\pi \hbar l_a}{m_\chi^3}\epsilon_\chi^2.
	\end{equation}
	\subsection{Selected Neutron star EoSs}
	To model the neutron star one must also have an EoS for the nuclear matter at extreme densities. Since it is not known with accuracy how matter behaves at extreme densities, whether particles at extreme densities change to hyperons or make deconfined quark matter which is known as strange matter. For baryonic matter we employees EoSs developed based on the quark meson coupling model \cite{RIKOVSKASTONE2007341}. Namely, for the case of  nucleons only or nucleon plus hyperons, N-QMC700 and F-QMC700 have been considered, while for strange matter the MIT bag model has been used \cite{Urbanec_2013}. In this way three models, neutron stars made of nucleons only, neutron stars made of nucleons and hyperons at higher densities, and neutron stars having nucleons and strange matter at higher densities may be studied.  \\
	The parameterized pressure and energy density relations (EoSs) for the QMC model are given below. The suitable values of the parameters for N-QMC700 and F-QMC700 are given in Table 1. 
	\begin{equation}
	P = \frac{N_1\epsilon^{p1}}{1+e^{(\epsilon - r)/a}}  + \frac{N_2\epsilon^{p2}}{1+e^{-(\epsilon - r)/a}} .
	\end{equation}
	This fit works well for the energy density up to 1200 MeV/fm$^3$ where $\epsilon$ and P are the energy density and pressure of the ordinary matter (baryonic matter), respectively and values for the constants $N_1$, $N_2$, $p_1$, $p_2$, '$r$' and '$a$' are given in Table 1.
	\begin{center}
		\begin{table}[h]
			\begin{tabular}{|c|c|c|c|c|c|c|}
				\hline
				& $N_1$ & $p_1$ & $N_2$ & $p_2$ & $r$ & $a$ \\ [0.5ex] 
				\hline\hline
				N-QMC700 &   0  & 0 &0.008623& 1.548& 342.4 &184.4\\
				\hline
				F-QMC-700 &0.0000002.62& 3.197& 0.0251& 1.286& 522.1 & 113\\
				\hline
			\end{tabular}
			\caption{Table for parameters taken from Ref. \cite{RIKOVSKASTONE2007341}}
		\end{table}
	\end{center}
	
	The strange matter EoS state is taken from Ref. \cite{Urbanec_2013}, where it was suggested that the nuclear matter may form a deconfined quark matter or strange matter at the core as the energy density increases which is also suggested in Refs. \cite{Terazawa:2001gg,PhysRevD.30.2379,Weber_1999,Glendenning:1997tb,Husain_2021}.
	The EoS of a neutron star with strange matter is given by
	\begin{equation}
	P = \frac{1}{3}(\epsilon - 4B),
	\end{equation}
	where B is the bag constant, with its value taken to be $10^{14}$gm/cm$^3$, and $\epsilon$ and P are the energy density and pressure of the strange matter, respectively. 
	\subsection{Structural Equations: Modified TOV Equation For Two Fluids}
	There is a continuous distribution of two fluids, dark matter and ordinary matter, inside the neutron stars. Such stars are gravitationally formed by the gravitational fields coupled to both types of matter, the TOV equation for a single type of matter \cite{Tolman169,PhysRev.55.374} is not sufficient. An energy-momentum tensor for such mixed fluid is solved leading to the following modified TOV equation for two fluids \cite{CIARCELLUTI201119,Sandin_2009}.
	\begin{equation}
	\frac{dP_{nucl}}{dr} = - \frac{[\epsilon_{nucl}(r)+P_{nucl}(r)][4\pi r^3(P_{nucl}(r) +  P_{DM}(r))+m(r)]}{r^2(1-\frac{2m(r)}{r})},
	\end{equation}
	\begin{equation}
	m_{nucl}(r) = 4\pi\int_{0}^{r}dr.r^2 \epsilon_{nucl}(r) ,
	\end{equation}
	\begin{equation}
	\frac{dP_{DM}}{dr} = - \frac{[\epsilon_{DM}(r)+P_{DM}(r)][4\pi r^3(P_{nucl}(r) +  P_{DM}(r))+m(r)]}{r^2(1-\frac{2m(r)}{r})} ,
	\end{equation}
	\begin{equation}
	m_{DM}(r) = 4\pi\int_{0}^{r}dr.r^2 \epsilon_{DM}(r),
	\end{equation}
	\begin{equation}
	m(r) = m_{nucl}(r) + m_{DM}(r).
	\end{equation}
	Equations (2.8) to (2.12) have been integrated together with the DM EoSs (fermionic and bosonic) and ordinary matter (N-QMC700, F-QMC700 and strange matter) EoSs, and the following results have been obtained. For calculating the tidal deformability the method suggested in \cite{Hinderer_2008,Hinderer_2010} have been adopted with modifications needed to apply for two fluids neutron stars.
	\section{Results}
	In this article the DM particle mass $m_\chi $ = 1 GeV is considered. As suggested in Refs. \cite{Li_2012,Kouvaris_2012,Ellis_2018,Mukhopadhyay_2017} the self-interacting scattering length $l_\chi$ = 1 fm for bosonic DM, and dark photon mass (self-interacting exchange particle mass)  $m_I$ = 100 MeV for fermionic DM are selected. 
	The relatively large dark matter fractions considered here do not lead to core collapse of the neutron star and they still satisfy the maximum mass constraint. This is because the QMC model-based equation of state is stiff enough to accommodate these dark matter contributions. 
	As suggested in Ref.\cite{2000NuPhB.564..185M} the low mass of the DM particles considered here, along with their repulsive self-interaction, leads to a large Chandrasekhar mass and prevents the DM halo from collapse.
	Similar conclusions have been reached with comparable dark matter content by a number of authors  \cite{Li_2012,Das_2020,CIARCELLUTI201119,Ellis_2018,Sandin_2009,Leung:2011zz}. Noticeably in reference \cite{CIARCELLUTI201119} different DM mass percentages up to 70\% of total mass are considered and it is shown that the neutron stars can contain DM with a mass of more than 15\% (approximately) of the total neutron star mass and still give a neutron star mass of 2 solar masses. They suggest that with the increment of DM contribution in the neutron star the maximum mass of neutron star decreases till DM contributes the 50\% of the total neutron star mass and at this ratio maximum mass of neutron star is approximately 1.7 solar masses. If the DM contribution increases further the maximum mass of neutron star increases with it. In our work, within the models considered, we investigate whether core collapse happens when neutron stars have certain ratio of DM mass to the total neutron star mass and whether they can still produce maximum mass of 2M, maintaining the ratio of DM mass in the core.
	
	\par In the following figures the central energy density for nuclear matter is kept fixed and the DM contribution is increased inside the neutron star.    
	\begin{figure}
		\centering 
		\includegraphics[width=.55\textwidth]{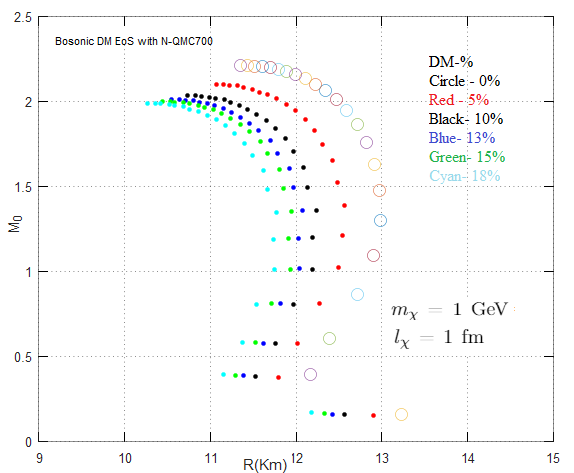}
		\caption{Total mass vs radius for the N-QMC700 EoS (contains nuclear matter only) with bosonic DM. For $l_\chi$ = 1 fm and $m_\chi$ = 1 GeV, different contributions of the DM mass to the total neutron star mass are considered.}
	\end{figure}
	\begin{figure}
		\centering 
		\includegraphics[width=.55\textwidth]{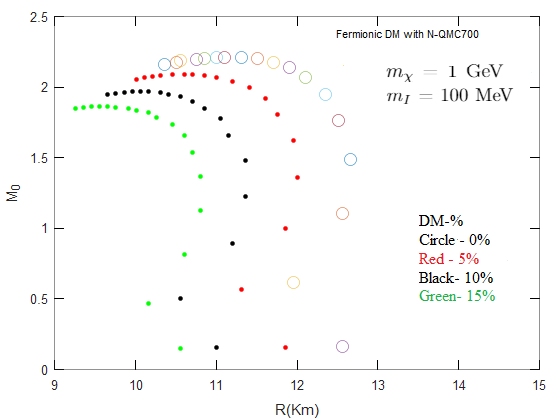}
		\caption{Total mass vs radius for the N-QMC700 EoS (contains nuclear matter only) with fermionic DM. For $m_\chi$ = 1 GeV and  $m_I$ = 100 MeV, different  contributions of the DM mass to the total neutron star mass are considered.}
	\end{figure}
	Both Figs. 1 and 2 are constructed assuming that the nuclear matter consists only of nucleons. Figure 1 is constructed using the N-QMC700 EoS and bosonic DM, while Figure 2 is constructed for the N-QMC700 EoS with fermionic DM. Figures 1 and 2 show the mass and radius relationship of the neutron stars for different contributions of bosonic and fermionic DM to the total mass of the neutron stars. As more dark matter is accreted by the neutron star, the total mass and the radius of the neutron star decreases. The observational constraint for the total (maximum) neutron star mass suggests that a neutron star should have a maximum mass of two solar masses. This constraint is satisfied when the ratio of DM mass to the total neutron star mass lies below 15\%. As the DM mass increases further the maximum mass of the neutron star moves below 2 solar masses. There is a significant difference in the radii of bosonic and fermionic DM neutron stars when 15\% of the total neutron star mass comes from DM. The neutron stars with N-QMC700 and fermionic DM at the core are relatively lighter and smaller in size in comparison to their bosonic counterparts.  \par
	Figures 3 and 4 are constructed for neutron stars which contain hyperons at the higher energy density of ordinary matter with DM at the core. Figure 3 corresponds to F-QMC700 with bosonic DM at the core, while Figure 4 presents total mass vs radius for F-QMC700 with fermionic DM. As the neutron stars accrete DM particles their capacity to have more mass decreases, as does the corresponding radius of the neutron stars. When hyperons are included in the neutron stars the maximum mass of the neutron stars become smaller than 2 solar masses when DM accretion reaches 5\%. The sizes of F-QMC700 neutron stars with bosonic and fermionic DM inside the core, are significantly different for the same percentage of DM mass inside the core. 
	\begin{figure}
		\centering 
		\includegraphics[width=.55\textwidth]{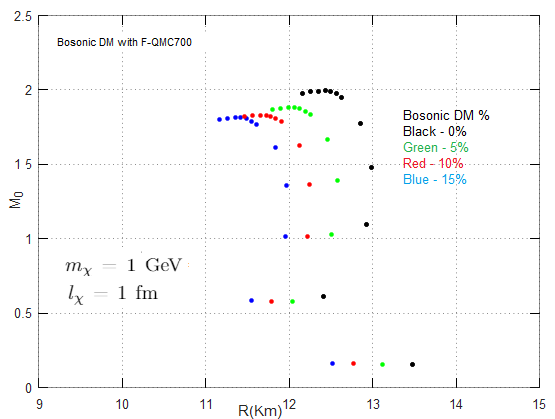}
		\caption{Total mass vs radius for F-QMC700 EoS (includes hyperons above an energy density of 600 MeV/fm$^3$) with bosonic DM. For $l_\chi$ = 1 fm and $m_\chi$ = 1 GeV, different \% of contribution of the DM mass to the total neutron star mass are considered.}
	\end{figure}
	\begin{figure}
		\centering 
		\includegraphics[width=.55\textwidth]{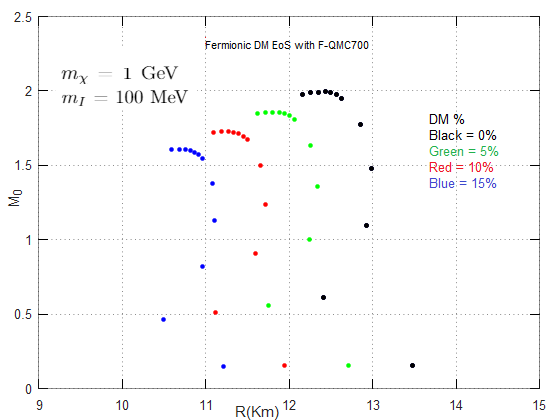}
		\caption{Total mass vs radius for F-QMC700 EoS (includes hyperons at energy density 600 MeV/fm$^3$) with fermionic DM EoS. For $m_\chi$ = 1 GeV and  $m_I$ = 100 MeV, different \% of contribution of the DM mass to the total neutron star mass is considered.}
	\end{figure}
	\par Figures 5 and 6 show the relation between mass and radius of neutron stars which contain strange matter at the core with DM. Figure 5 corresponds to strange matter (deconfined quark matter) EoS with bosonic DM and Figure 6 shows the results for a strange matter EoS with fermionic DM. Similar to F-QMC700 neutron stars, strange matter EoS neutron stars with DM have a maximum mass of less than 2M$_\odot$ when DM particles contribute just 5\% of the total mass of the neutron stars. There is not a significant difference in the total mass and radius of strange matter neutron stars at various stages of DM accretion of bosonic and fermionic DM as for F-QMC700.
	\begin{figure}
		\centering 
		\includegraphics[width=.55\textwidth]{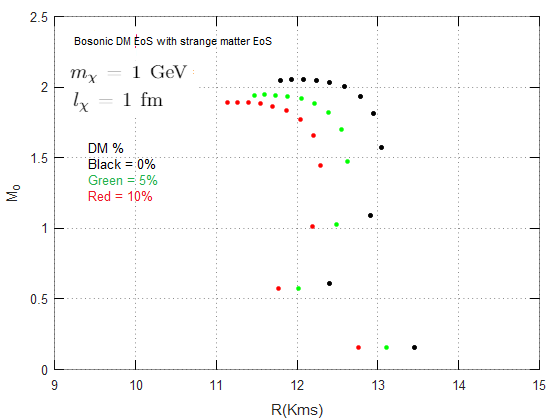}
		\caption{Total mass vs radius for strange matter EoS with bosonic DM EoS. For $l_\chi$ = 1 fm and $m_\chi$ = 1 GeV, different contributions of the DM mass to the total neutron star mass are considered.}
	\end{figure}
	\begin{figure}
		\centering 
		\includegraphics[width=.55\textwidth]{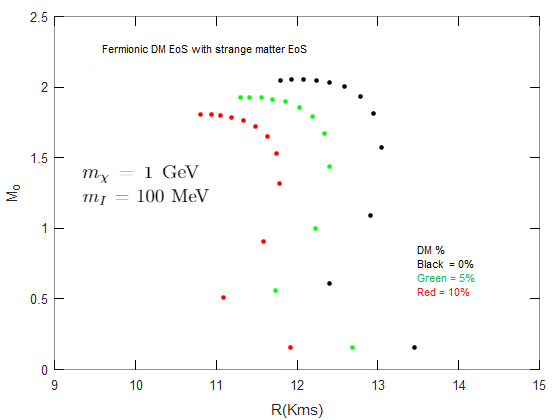}
		\caption{Total mass vs radius for strange matter EoS with fermionic DM EoS. For $m_\chi$ = 1 GeV and  $m_I$ = 100 MeV, different of contributions of the DM mass to the total neutron star mass are considered.}
	\end{figure}
	\par From Figures 7 to 12 we can obtain some insight \cite{Nelson:2018xtr} into the structure of neutron stars containing DM which contributes 5\% of the total mass. These figures give snapshots of the distribution of energy densities inside the neutron star, from the centre to the surface.  Figures 7, 9 and 11 represent the distribution of bosonic DM energy density and  F-QMC700 EoS, strange matter EoS and N-QMC700 EoS energy density, respectively, while  Figures 8, 10 and 12 display the distribution of fermionic DM energy density and F-QMC700 EoS, strange matter EoS and N-QMC700 EoS energy density, from the centre towards the surface. In these three cases bosonic DM condenses inside the core and remains inside the surface of the neutron star within a radius of a few kilometers. In contrast fermionic DM does not condense and covers the whole neutron star from the core to outside the surface, because the  fermionic DM pressure does not vanish before the nuclear matter pressure as it is shown in the Figs. 8, 10 and 12. For instance the distances form the centre of the neutron star to the points where the fermionic DM vanishes are N-QMC700 has radius 220.3 km, the F-QMC700 has radius 223.47 km while strange matter has the largest radius 225.9 km. A similar phenomenon, namely DM a DM halo around the neutron star, was reported in Ref. \cite{Nelson:2018xtr} Compared to bosonic DM, fermionic DM requires a smaller energy density at the centre to contribute 5\% of the total mass of the neutron star. 
	\begin{figure}
		\centering 
		\includegraphics[width=.55\textwidth]{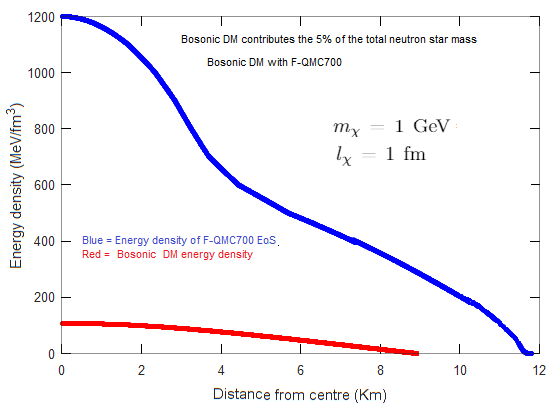}
		\caption{Hadronic and bosonic DM energy density distribution inside the neutron star which contains 5\% of DM mass to the total neutron star mass for F-QMC700.}
	\end{figure}
	\begin{figure}
		\centering 
		\includegraphics[width=.55\textwidth]{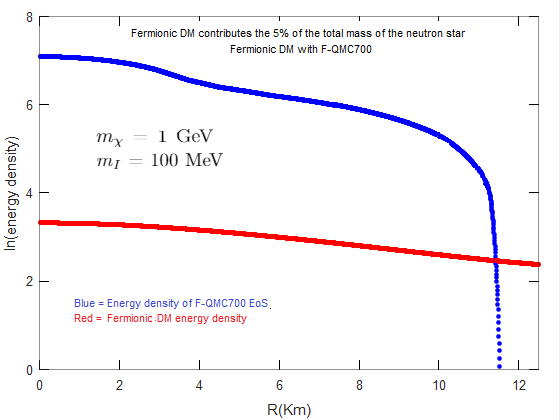}
		\caption{Hadronic and fermionic DM energy density distribution inside the neutron star which contains 5\% of DM mass to the total neutron star mass for F-QMC700.}
	\end{figure}
	\begin{figure}
		\centering 
		\includegraphics[width=.55\textwidth]{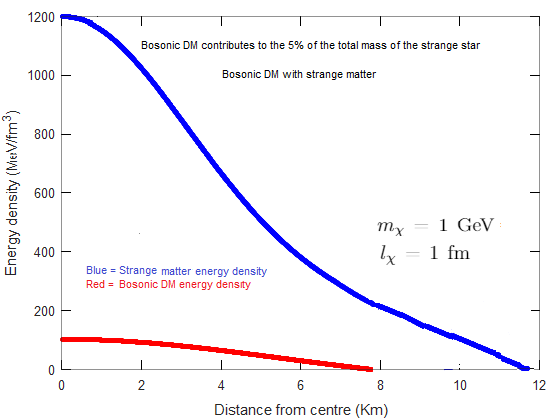}
		\caption{Distribution of the energy density of strange matter and bosonic DM energy density inside the neutron star when bosonic DM contributes 5\% of the total mass of the neutron star.}
	\end{figure}
	\begin{figure}
		\centering 
		\includegraphics[width=.55\textwidth]{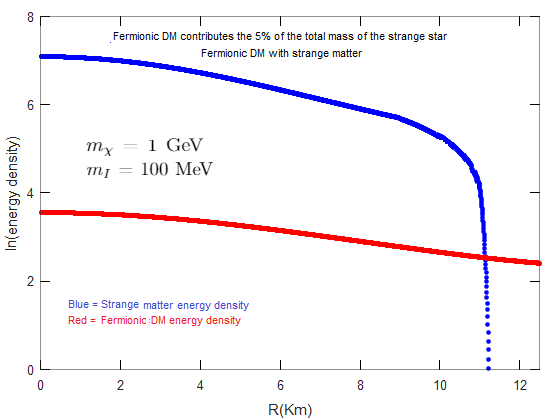}
		\caption{Distribution of the energy density of strange matter and fermionic DM energy density inside the neutron star when fermionic DM contributes 5\% of the total mass of the neutron star.}
	\end{figure}
	\begin{figure}
		\centering 
		\includegraphics[width=.55\textwidth]{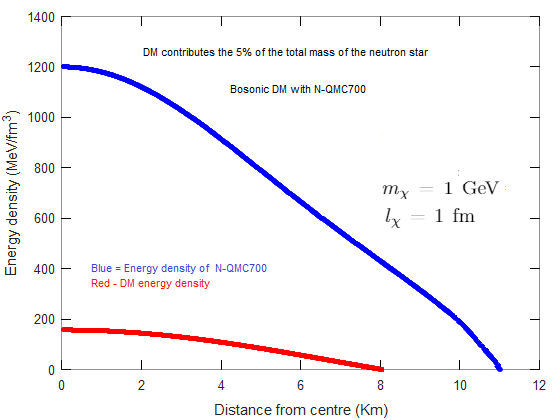}
		\caption{Distribution of nuclear matter (N-QMC700) and bosonic DM energy density inside the neutron star when bosonic DM contributes 5\% of the total mass of the neutron star.}
	\end{figure}
	\begin{figure}
		\centering 
		\includegraphics[width=.55\textwidth]{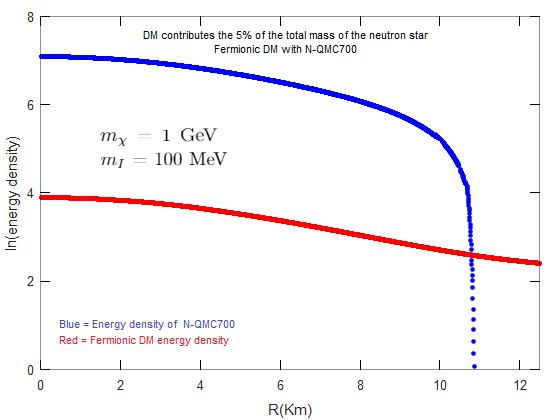}
		\caption{Distribution of nuclear matter (N-QMC700) and fermionic DM energy density inside the neutron star when fermionic DM contributes 5\% of the total mass of the neutron star.}
	\end{figure}
	\begin{figure}
		\centering 
		\includegraphics[width=.55\textwidth]{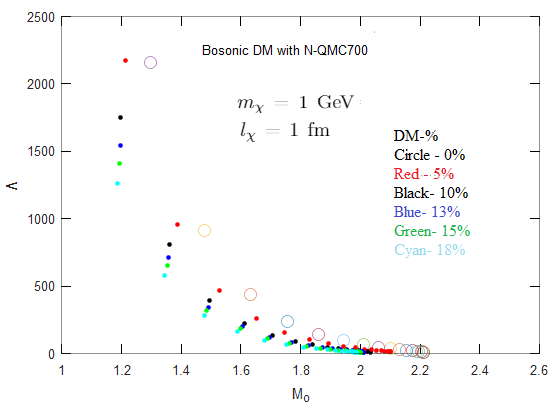}
		\caption{Change in tidal deformability of a neutron star consist nuclear matter (N-QMC700) with a different contributions of bosonic DM mass to the total neutron star mass.}
	\end{figure}
	\par
	In Figure 13  the tidal deformability of the neutron stars is shown against the total mass of the neutron stars. As the DM increases inside the neutron star the tidal deformability of the neutron star decreases, because the neutron star becomes more compact. QMC700 neutron stars have higher and strange matter neutron stars have lower values of tidal deformability when we do not consider DM inside them while QMC700 and strange matter satisfy the tidal deformability constraint if they contain certain amount of DM inside them. The DM contribution to neutron star mass that satisfy the tidal deformability constraint is as follows. For a neutron star of mass 1.4 M$_\odot$ having a DM mass between 5\% to 18\% of the total mass satisfies the tidal deformability constraint imposed by gravitational wave detection  %\cite{Abbott_2017,Abbott_2019}
	. As shown in Figure 14, for neutron stars with F-QMC700 and bosonic DM the tidal deformability decreases with increasing DM content inside the neutron star. The bosonic DM contribution between 5\% to 15\% for a neutron star of 1.4M$_\odot$ satisfies the the tidal deformability constraint. Figure 15 shows a neutron star having strange matter at the core with bosonic DM. The tidal deformability constraint is followed by a neutron star of mass 1.4M$_\odot$ only when bosonic DM mass contribution to the total mass is less than 5\%.  
	\par Figure 16, 17 and 18 are plotted for fermionic DM with N-QMC700, F-QMC700 and strange matter EoSs, respectively. Figure 16 suggests that for a neutron star of mass 1.4M$_\odot$  the constraint on tidal deformability is only satisfied when fermionic DM contribution to the total mass is between 5\% to 10\% while for bosonic DM case it is satisfied up to 18\%. Figure 17 indicates that a neutron star of mass 1.4M$_\odot$  the constraint on tidal deformability is satisfied when fermionic DM contribution to the total mass is between 5\% to 10\%, while bosonic DM it is satisfied up to 15\%. As suggested in Figure 18, a neutron star of mass 1.4M$_\odot$  the constraint on tidal deformability only is followed when fermionic DM contribution to the total mass is less than 5\%.
	The properties of neutron stars are quite different with bosonic and fermionic DM particularly the distribution of DM energy density. 
	\begin{figure}
		\centering 
		\includegraphics[width=.55\textwidth]{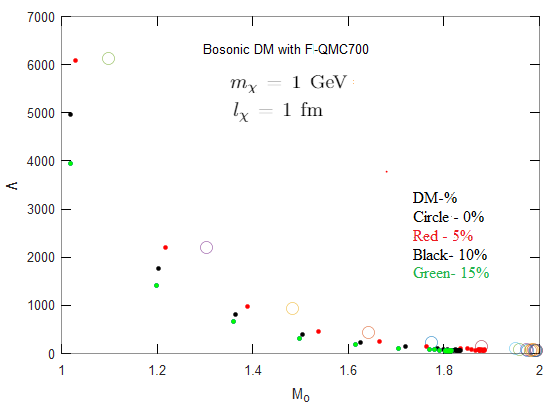}
		\caption{Change in tidal deformability of neutron star having nuclear matter including hyperons (F-QMC700) with different contributions of bosonic DM mass to the total neutron star mass.}
	\end{figure}
	\begin{figure}
		\centering 
		\includegraphics[width=.55\textwidth]{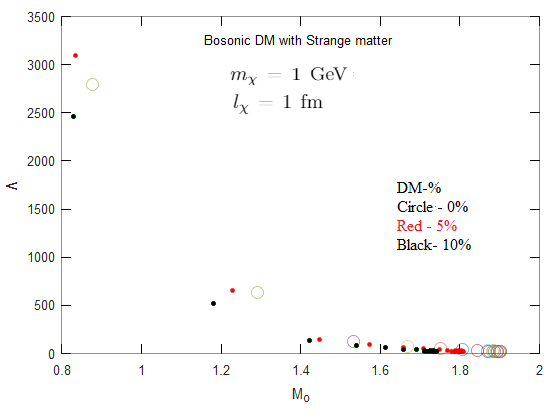}
		\caption{Change in tidal deformability of neutron star having strange matter with different contributions of bosonic DM mass to the total neutron star mass.}
	\end{figure}
	\begin{figure}
		\centering 
		\includegraphics[width=.55\textwidth]{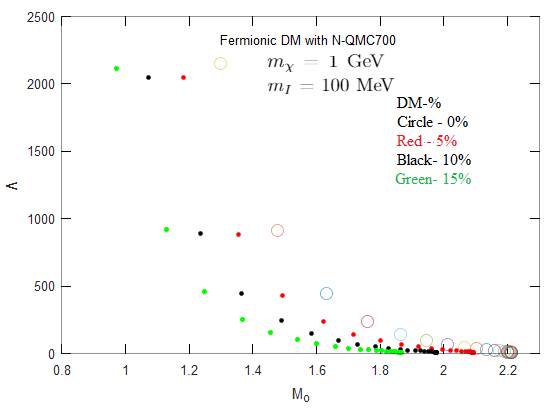}
		\caption{Change in tidal deformability of a neutron star having nuclear matter with different contributions of fermionic DM mass to the total neutron star mass.}
	\end{figure}
	\begin{figure}
		\centering 
		\includegraphics[width=.55\textwidth]{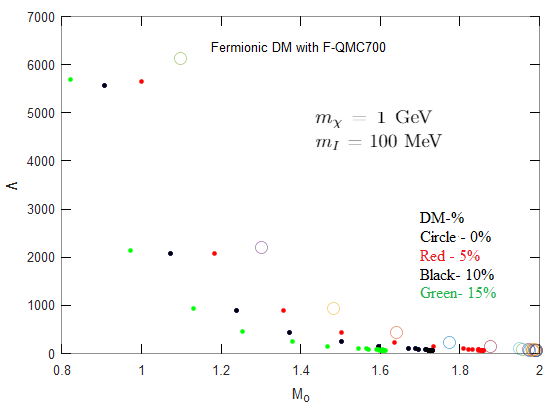}
		\caption{Change in tidal deformability of a neutron star having nuclear matter including hyperons with different contributions of fermionic DM mass to the total neutron star mass.}
	\end{figure}
	\begin{figure}[h]
		\centering 
		\includegraphics[width=.55\textwidth]{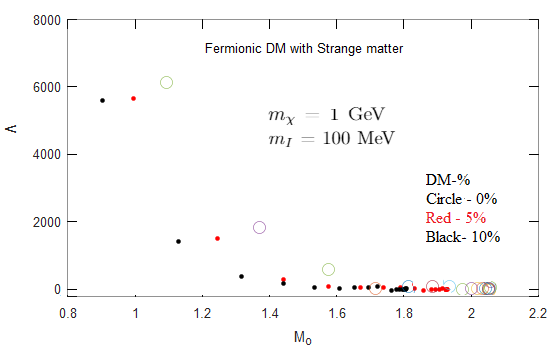}
		\caption{Change in tidal deformability of a neutron star having strange matter with different contributions of fermionic DM mass to the total neutron star mass.}
	\end{figure}
	\newpage
	\section{Conclusion}
	A number of studies of DM accretion on a neutron star suggest that it may take several trillion or even more number of years \cite{Li_2012,bell2020nucleon,PhysRevD.82.063531} for a neutron star to accumulate sufficient DM matter that it constitutes 5\% of the total mass. Particularly in Ref. \cite{bell2020nucleon}  different kind of DM interactions are considered, and they have suggested that capture rate is dependent on the type on interactions.
	Frankly, our knowledge of dark matter is so poor that it is difficult to say what is or is not a likely scenario.% For example, most calculations which model the capture of DM, ignore the effect of self-interactions of the dark matter, which can be much stronger than the interactions with regular matter. As in the lower capture rate cases much of the dark matter passes through the star without interaction, the situation could change drastically as dark matter starts to build up.
	We do not take a position on the likelihood of reaching the fractions considered here. Rather we explore the consequences if they do.
	We also note that there is a different dark matter scenario where the fractions of dark matter considered here would indeed be possible. A few years ago Fornal and Grinstein \cite{Fornal20181,Fornal_2020} proposed an explanation for the anomaly in the neutron lifetime, whereby with a small branching ratio the neutron might decay to a dark matter particle almost degenerate with the neutron. While neutron star properties rule this out for inert dark matter, one can obtain neutron stars within the observed mass range provided that the dark matter has a fairly strong repulsive self-interaction. In that case there can be a fraction of the star which is dark matter, comparable with the fractions considered here \cite{Motta:2018bil,Motta:2018rxp,refId0}.
\par
	Having 5\% of the mass contributed by DM is a large amount of mass, in the sense that it leads to significant changes in the properties of the neutron stars. Having 5\% or less DM mass in the core seems a good and realistic approximation to see the changes of the neutron star properties. From mass vs radius plots it is evident that in general having DM in the core will reduce the radius and maximum mass of the neutron star. Bosonic DM neutron stars may be heavier and bigger than fermionic DM neutron stars. \\
	We have shown a comparative study of three models of neutron stars with DM of dark matter particle mass, m$_\chi$ = 1 GeV (for bosonic DM self-scattering length l$_\chi$= 1 fm and for fermionic DM self-interacting particle mass m$_I$ = 100 MeV). The strange matter EoS produces neutron stars with a maximum mass very close to 2M$_\odot$ when 5\% of the mass is contributed by DM. F-QMC700 is also very close to producing a maximum mass of 2M$_\odot$ for the neutron star but strange matter EoS is closer. The N-QMC700 EoS can produce neutron stars with a maximum mass of more than 2 solar masses, even when 15\% of the mass is  contributed by DM. For selected values of the DM particle properties and type of interaction, the constraint of maximum mass of at least 2 solar masses  and radius in the range 9 to 13 kms stars is respected by all three types of EoSs. \\
	The distribution of the energy density of DM inside the neutron stars \cite{Nelson:2018xtr} suggests that fermionic DM covers the whole neutron star right from the centre to the surface and outside, where it constitutes a halo while bosonic DM sits inside the surface of the neutron star.\\ 
	In light of constraint on tidal deformability, in general the neutron stars can possess greater amount of bosonic DM than fermionic DM. As the fermionic DM contribution can be at most 10\% of the total neutron star mass for N-QMC700 and F-QMC700 while it can 18\% for bosonic DM with N-QMC700 for have a tidal deformability in the range $\Lambda =$ 400-800. For a strange matter neutron star the tidal deformability constraint suggests that whether it is fermionic DM or bosonic DM, the DM mass inside the neutron star can be no more than 5\% of the total neutron star mass. As shown in this article the distribution of energy density of fermionic and bosonic DM is very different. This study explores the effects of different kinds of DM (with certain properties) on neutron stars and compares with the observational constraints. The effects of Bosonic DM matter are found to be quite different from the effects of Fermionic DM.  Our study suggests constraints on how much dark matter (with assumed properties) neutron stars, made of different kinds of matter, may contain while remaining consistent with the observational constraint on mass, radius, and tidal deformability. As the sensitivity of the gravitational wave detectors grows over the next few years one may expect these constraints to become much more stringent. 
		We note that the capture of DM on a neutron star may have various dramatic consequences apart from the collapse of neutron stars to black holes \cite{PhysRevD.40.3221,Garani_2019_30}. It may modify the gravitational wave signatures from binary neutron star mergers \cite{Nelson:2018xtr,Ellis:2017jgp,Bramante_2018_31}, or the DM halo reported here, which extends beyond the star, may have its own observational consequences. 
	 \par In future it will be interesting to see the gravitational lensing around the neutron star containing DM inside it, that may suggest the nature of DM because gravitational lensing will be different for different distributions of energy densities.
	\newpage
	%\begin{table}[tbp]
	%\centering
	%\begin{tabular}{|lr|c|}
	%\hline
	%x&y&x and y\\
	%\hline
	%a & b & a and b\\
	%1 & 2 & 1 and 2\\
	%$\alpha$ & $\beta$ & $\alpha$ and $\beta$\\
	%\hline
	%\end{tabular}
	%\caption{\label{tab:i} We prefer to have borders around the %tables.}
	%\end{table}
	
	\acknowledgments
	This research was supported by a University of Adelaide International Scholarship, the ARC Centre of Excellence for Dark Matter Particle Physics and the Centre for the Subatomic Structure of Matter (CE Z00100008) and an ARC Discovery Project (DF 180100497). 
	%This is the most common positions for acknowledgments. A macro is
	%available to maintain the same layout and spelling of the heading.
	
	%\paragraph{Note added.} This is also a good position for notes added %\cite{Mukhopadhyay_2017}
	%after the paper has been written.

	% in short all the information that clearly identify a document.

	%\begin{thebibliography}{99}
	
	%\bibliography{main.bbl}

\begin{thebibliography}{99}
		
		\bibitem{Mukhopadhyay_2017}
		Mukhopadhyay, Somnath and Atta, Debasis and Imam, Kouser and Basu, D. N. and Samanta, C., \emph{Compact bifluid hybrid stars: hadronic matter mixed with self-interacting fermionic asymmetric dark matter}, \emph{The European Physical Journal C} {\bf 77} (2017) 7.
		
		\bibitem{1978AJ.....83.1050S}
		{Steigman}, G. and {Sarazin}, C.~L. and {Quintana}, H. and {Faulkner}, J., \emph{Dynamical interactions and astrophysical effects of stable heavy neutrinos.},
		\emph{aj} {\bf 83} (1978) 1050-1061.
		
		\bibitem{1985ApJ...294..663S}
		{Spergel}, D.~N. and {Press}, W.~H., \emph{Effect of hypothetical, weakly interacting, massive particles on energy transport in the solar interior}, \emph{apj} {\bf 294} (1985) 663-673.
		
		\bibitem{1985ApJ...299..994F}
		{Faulkner}, J. and {Gilliland}, R.~L., \emph{Weakly interacting, massive particles and the solar neutrino flux},
		\emph{apj} {\bf 299} (1985) 994-1000.
		
		\bibitem{PhysRevD.77.043515}
		Bertone, Gianfranco and Fairbairn, Malcolm, \emph{Compact stars as dark matter probes}, \emph{Phys. Rev. D} {\bf 77} (2008) 9.
		
		\bibitem{PhysRevD.77.023006}
		Kouvaris, Chris, \emph{WIMP annihilation and cooling of neutron stars},
		\emph{Phys. Rev. D} {\bf 77} (2008) 9.
		
		\bibitem{PhysRevD.40.3221}
		Goldman, Itzhak and Nussinov, Shmuel, \emph{Weakly interacting massive particles and neutron stars}, \emph{Phys. Rev. D} {\bf 40} (1989) 3221--3230.
		
		\bibitem{f2b55bdf3fc1466081f4c0d091544a2d}
		Andrew Gould and Draine, {Bruce T.} and Romani, {Roger W.} and Shmuel Nussinov, \emph{Neuton stars: Graveyard of charged dark matter},
		\emph{Physics Letters, Section B: Nuclear, Elementary Particle and High-Energy Physics} {\bf 238} (1990) 337--343.
		
		\bibitem{PhysRevLett.105.011301}
		Frandsen, Mads T. and Sarkar, Subir, \emph{Asymmetric Dark Matter and the Sun}, \emph{Phys. Rev. Lett.} {\bf 105} (2010) 011301.
		
		\bibitem{Lopes_2012}
		Il{\'{\i}}dio Lopes and Joseph Silk, \emph{{SOLAR} {NEUTRINO} {PHYSICS}: {SENSITIVITY} {TO} {LIGHT} {DARK} {MATTER} {PARTICLES}}, \emph{American Astronomical Society} {\bf 752} (2012) 2.
		
		\bibitem{Taoso_2010}
		Taoso, Marco and Iocco, Fabio and Meynet, Georges and Bertone, Gianfranco and Eggenberger, Patrick, \emph{Effect of low mass dark matter particles on the Sun}, \emph{Physical Review D} {\bf 82} (2010) 8.
		
		\bibitem{CIARCELLUTI201119}
		Paolo Ciarcelluti and Fredrik Sandin, \emph{Have neutron stars a dark matter core?}, \emph{Physics Letters B} {\bf 695} (2011) 19-21.
		
		\bibitem{Sandin_2009}
		Sandin, Fredrik and Ciarcelluti, Paolo, \emph{Effects of mirror dark matter on neutron stars}, \emph{Astroparticle Physics} {\bf 32} (2009) 278-284.
		
		\bibitem{Leung:2011zz}
		Leung, S. C. and Chu, M. C. and Lin, L. M., \emph{Dark-matter admixed neutron stars}, \emph{Phys. Rev. D} {\bf 84} (2011) 107301.
		
		\bibitem{B_hmer_2007}
		Böhmer, C G and Harko, T, \emph{Can dark matter be a Bose–Einstein condensate?}, \emph{Journal of Cosmology and Astroparticle Physics} {\bf 2007} (2007) 025–025.
		
		\bibitem{Rocha_2013}
		Rocha, Miguel and Peter, Annika H. G. and Bullock, James S. and Kaplinghat, Manoj and Garrison-Kimmel, Shea and Oñorbe, Jose and Moustakas, Leonidas A., \emph{Cosmological simulations with self-interacting dark matter – I. Constant-density cores and substructure}, \emph{Monthly Notices of the Royal Astronomical Society} {\bf 430} (2013) 81-104.
		
		\bibitem{Zavala_2013}
		Zavala, Jesús and Vogelsberger, Mark and Walker, Matthew G., \emph{Constraining self-interacting dark matter with the Milky Way’s dwarf spheroidals}, \emph{Monthly Notices of the Royal Astronomical Society: Letters} {\bf 431} (2013) L20-L24.
		
		\bibitem{Peter_2013}
		Peter, Annika H. G. and Rocha, Miguel and Bullock, James S. and Kaplinghat, Manoj, \emph{Cosmological simulations with self-interacting dark matter – II. Halo shapes versus observations}, \emph{Monthly Notices of the Royal Astronomical Society} {\bf 430} (2013) 105-120.
		
		\bibitem{Spergel_2000}
		Spergel, David N. and Steinhardt, Paul J., \emph{Observational Evidence for Self-Interacting Cold Dark Matter}, \emph{Physical Review Letters} {\bf 84} (2000) 3760–3763.
		
		\bibitem{Faraggi_2002}
		Faraggi, Alon E. and Pospelov, Maxim, \emph{Self-interacting dark matter from the hidden heterotic-string sector}, \emph{Astroparticle Physics} {\bf 16} (2002) 451–461.
		
		\bibitem{Wandelt:2000ad}
		Wandelt, Benjamin D. and Dave, Romeel and Farrar, Glennys R. and McGuire, Patrick C. and Spergel, David N. and Steinhardt, Paul J., \emph{Selfinteracting dark matter}, \emph{arXiv } {\bf astro-ph/0006344} (2000) 263--274.
		
		\bibitem{Mohapatra_2002}
		Mohapatra, R. N. and Nussinov, S. and Teplitz, V. L., \emph{Mirror matter as self-interacting dark matter}, \emph{Physical Review D} {\bf 66} (2002) 6.
		
		\bibitem{Kusenko_2001}
		Kusenko, Alexander and Steinhardt, Paul J., \emph{Q-Ball Candidates for Self-Interacting Dark Matter}, \emph{Physical Review Letters} {\bf 87} (2001) 14.
		
		\bibitem{Loeb_2011}
		Loeb, Abraham and Weiner, Neal, \emph{Cores in Dwarf Galaxies from Dark Matter with a Yukawa Potential}, \emph{Physical Review Letters} {\bf 106} (2011) 17.
		
		\bibitem{Kouvaris_2012}
		Kouvaris, Chris, \emph{Limits on Self-Interacting Dark Matter from Neutron Stars}, \emph{Physical Review Letters} {\bf 108} (2012) 19.
		
		\bibitem{2013MNRAS.430.1722V}
		{Vogelsberger}, Mark and {Zavala}, Jesus, \emph{Direct detection of self-interacting dark matter}, \emph{mnras} {\bf 430} (2013) 3.
		
		\bibitem{Kaplinghat_2014}
		Kaplinghat, Manoj and Keeley, Ryan E. and Linden, Tim and Yu, Hai-Bo, \emph{Tying Dark Matter to Baryons with Self-Interactions}, \emph{Physical Review Letters} {\bf 113} (2014) 2.
		
		\bibitem{Cline_20141}
		Cline, James M. and Liu, Zuowei and Moore, Guy D. and Xue, Wei, \emph{Scattering properties of dark atoms and molecules}, \emph{Physical Review D} {\bf 89} (2014) 4.
		
		\bibitem{Cline_2014}
		Cline, James M. and Liu, Zuowei and Moore, Guy D. and Xue, Wei, \emph{Composite strongly interacting dark matter}, \emph{Physical Review D} {\bf 90} (2014) 90.
		
		\bibitem{Petraki_2014}
		Petraki, Kalliopi and Pearce, Lauren and Kusenko, Alexander, \emph{Self-interacting asymmetric dark matter coupled to a light massive dark photon}, \emph{Journal of Cosmology and Astroparticle Physics} {\bf 2014} (2014) 07.
		
		%\bibitem{Boddy_2014}
		%Boddy, Kimberly K. and Feng, Jonathan L. and Kaplinghat, Manoj and Tait, Tim M. P., \emph {Self-interacting dark matter from a non-Abelian hidden sector}, \emph {Physical Review D} %{\bf 89} (2014) 11.$
		
		\bibitem{Schutz_2015}
		Schutz, Katelin and Slatyer, Tracy R., \emph{Self-scattering for Dark Matter with an excited state}, \emph{Journal of Cosmology and Astroparticle Physics} {\bf 2015} (2015) 021–021.
		
		\bibitem{Feng_2009}
		Feng, Jonathan L and Kaplinghat, Manoj and Tu, Huitzu and Yu, Hai-Bo, \emph{Hidden charged dark matter}, \emph{Journal of Cosmology and Astroparticle Physics} {\bf 2009} (2009) 004-004.
		
		\bibitem{Feng_2010}
		Feng, Jonathan L. and Kaplinghat, Manoj and Yu, Hai-Bo, \emph{Halo-Shape and Relic-Density Exclusions of Sommerfeld-Enhanced Dark Matter Explanations of Cosmic Ray Excesses}, \emph{Physical Review Letters} {\bf 104} (2010) 15.
		
		%\bibitem{Markevitch_2004}
		%Markevitch, M. and Gonzalez, A. H. and Clowe, D. and Vikhlinin, A. and Forman, W. and Jones, C. and Murray, S. and Tucker, W., \emph{ Direct Constraints on the Dark Matter %Self‐Interaction Cross Section from the Merging Galaxy Cluster 1E 0657−56}, \emph{ The Astrophysical Journal} {\bf 606} (2004) 819-824.
		
		\bibitem{Burgess_2001}
		Burgess, C.P. and Pospelov, Maxim and ter Veldhuis, Tonnis, \emph{The Minimal Model of nonbaryonic dark matter: a singlet scalar}, \emph{Nuclear Physics B} {\bf 619} (2001) 709-728.
		
		\bibitem{Kouvaris_2015}
		Kouvaris, Chris and Nielsen, Niklas Grønlund, \emph{Asymmetric dark matter stars}, \emph{Physical Review D} {\bf 92} (2015) 6.
		
		\bibitem{Ellis_2018}
		Ellis, John and Hütsi, Gert and Kannike, Kristjan and Marzola, Luca and Raidal, Martti and Vaskonen, Ville, \emph{Dark matter effects on neutron star properties}, \emph{Physical Review D} {\bf 97} (2018) 12.
		
		\bibitem{RIKOVSKASTONE2007341}
		J. {Rikovska Stone} and P.A.M. Guichon and H.H. Matevosyan and A.W. Thomas, \emph{Cold uniform matter and neutron stars in the quark–meson-coupling model}, \emph{Nuclear Physics A} {\bf 792} (2007) 341-369.
		
		\bibitem{BAYM1971225}
		Gordon Baym and Hans A. Bethe and Christopher J Pethick, \emph{Neutron star matter}, \emph{Nuclear Physics A} {\bf 175} (1971) 225 - 271.
		
		\bibitem{Tolman169}
		Tolman, Richard C., \emph{Effect of Inhomogeneity on Cosmological Models}, \emph{Proceedings of the National Academy of Sciences} {\bf 20} (1934) 169--176.
		
		\bibitem{PhysRevD.10.2599}
		Chodos, A. and Jaffe, R. L. and Johnson, K. and Thorn, C. B., \emph{Baryon structure in the bag theory}, \emph{Phys. Rev. D} {\bf 10} (1974) 2599--2604.
		
		\bibitem{Urbanec_2013}
		Urbanec, M and Miller, J. C and Stuchlík, Z, \emph{Quadrupole moments of rotating neutron stars and strange stars}, \emph{Monthly Notices of the Royal Astronomical Society} {\bf 433} (2013) 1903–1909.
		
		\bibitem{PhysRevLett.93.132502}
		Guichon, P. A. M. and Thomas, A. W., \emph{Quark Structure and Nuclear Effective Forces}, \emph{Phys. Rev. Lett.} {\bf 93} (2004) 132502.
		
		\bibitem{Husain_2020}
		Wasif Husain and Anthony W. Thomas, \emph{Significance of lower energy density region of neutron star and universalities among neutron star properties}, \emph{Journal of Physics: Conference Series} {\bf 1643} (2020) 012066.
		
		\bibitem{PhysRev.55.374}
		Oppenheimer, J. R. and Volkoff, G. M., \emph{On Massive Neutron Cores}, \emph{Phys. Rev.} {\bf 55} (1939) 374--381.
		
		\bibitem{Li_2012}
		Li, X.Y and Wang, F.Y and Cheng, K.S, \emph{Gravitational effects of condensate dark matter on compact stellar objects}, \emph{Journal of Cosmology and Astroparticle Physics} {\bf 2012} (2012) 031–031.
		
		\bibitem{L_2012}
		X.Y Li and T Harko and K.S Cheng, \emph{Condensate dark matter stars}, \emph{Journal of Cosmology and Astroparticle Physics} {\bf 2012} (2012) 001--001.
		
		\bibitem{Sandin_2009}
		Sandin, Fredrik and Ciarcelluti, Paolo, \emph{Effects of mirror dark matter on neutron stars}, \emph{Astroparticle Physics} {\bf 32} (2009) 278–284.
		
		\bibitem{CIARCELLUTI201119}
		Paolo Ciarcelluti and Fredrik Sandin, \emph{Have neutron stars a dark matter core?}, \emph{Physics Letters B} {\bf 695} (2011) 19-21.
		
		\bibitem{Nelson:2018xtr}
		Nelson, Ann and Reddy, Sanjay and Zhou, Dake, \emph{Dark halos around neutron stars and gravitational waves}, \emph{Journal of Cosmology and Astroparticle Physics} {\bf 07} (2019) 012.
		
		\bibitem{PhysRevD.81.123521}
		de Lavallaz, Arnaud and Fairbairn, Malcolm, \emph{Neutron stars as dark matter probes}, \emph{Phys. Rev. D} {\bf 81} (2010) 123521.
		
		\bibitem{PhysRevD.82.063531}
		Kouvaris, Chris and Tinyakov, Peter, \emph{Can neutron stars constrain dark matter?}, \emph{Phys. Rev. D} {\bf 82} (2010) 063531.
		
		\bibitem{Terazawa:2001gg}
		Terazawa, H., \emph{A new trend in high-energy physics: Current topics in nuclear and particle physics}, \emph{International Conference on New Trends in High-Energy Physics: Experiment, Phenomenology, Theory} {\bf KEK-PREPRINT-2001-129} (2001) 246--255.
		
		\bibitem{PhysRevD.30.2379}
		Farhi, Edward and Jaffe, R. L., \emph{Strange matter}, \emph{Phys. Rev. D} {\bf 30} (1984) 2379--2390.
		
		\bibitem{Weber_1999}
		F Weber, \emph{Quark matter in neutron stars}, \emph{Journal of Physics G: Nuclear and Particle Physics} {\bf 25} (1999) R195--R229.
		
		\bibitem{Glendenning:1997tb}
		Glendenning, Norman K., \emph{Strangeness in compact stars and signal of deconfinement}, \emph{J. Phys. G} {\bf 23} (1997) 2013--2027.
		
		\bibitem{Husain_2021}
		Husain, Wasif and Thomas, Anthony W., \emph{Hybrid stars with hyperons and strange quark matter}, \emph{PROCEEDINGS OF THE 14TH ASIA-PACIFIC PHYSICS CONFERENCE} {\bf http://dx.doi.org/10.1063/5.0036994} (2021).
		
		\bibitem{bell2020nucleon}
		Nicole F. Bell and Giorgio Busoni and Theo F. Motta and Sandra Robles and Anthony W. Thomas and Michael Virgato, \emph{Nucleon Structure and Strong Interactions in Dark Matter Capture in Neutron Stars}, \emph{arXiv} {\bf 2012.08918} (2020) 7.
		
		\bibitem{Hinderer_2008}
		Hinderer, Tanja, \emph{Tidal Love Numbers of Neutron Stars}, \emph{The Astrophysical Journal} {\bf 677} (2008) 1216–1220.
		
		\bibitem{Hinderer_2010}
		Hinderer, Tanja and Lackey, Benjamin D. and Lang, Ryan N. and Read, Jocelyn S., \emph{Tidal deformability of neutron stars with realistic equations of state and their gravitational wave signatures in binary inspiral}, \emph{Physical Review D} {\bf 81} (2010) 12.
		
		\bibitem{Flanagan_2008}
		Flanagan, Éanna É. and Hinderer, Tanja, \emph{Constraining neutron-star tidal Love numbers with gravitational-wave detectors}, \emph{Physical Review D} {\bf 77} (2008) 2.
		
		\bibitem{Das_2020}
		Flanagan, Éanna É. and Hinderer, Tanja,Das, H. C. and Kumar, Ankit and Kumar, Bharat and Kumar Biswal, Subrat and Nakatsukasa, Takashi and Li, Ang and Patra, S. K. \emph{Effects of dark matter on the nuclear and neutron star matter}, \emph{Mon. Not. Roy. Astron. Soc.} {\bf 495} (2020) 4893--4903.
		
		\bibitem{Garani_2019_30}
		Raghuveer Garani and Yoann Genolini and Thomas Hambye \emph{New analysis of neutron star constraints on asymmetric dark matter}, \emph{Journal of Cosmology and Astroparticle Physics} {\bf 2019} (2019) 035--035.
		
		\bibitem{Ellis:2017jgp}
		Ellis, John and Hektor, Andi and Hutsi, Gert and Kannike, Kristjan and Marzola, Luca and Raidal, Martti and Vaskonen, Ville \emph{Search for Dark Matter Effects on Gravitational Signals from Neutron Star Mergers}, \emph{Phys. Lett. B} {\bf 781} (2018) 607--610.
		
		\bibitem{Bramante_2018_31}
		Bramante, Joseph and Linden, Tim and Tsai, Yu-Dai \emph{Searching for dark matter with neutron star mergers and quiet kilonovae}, \emph{Phys. Lett. D} {\bf 97} (2018) 055016.
		
		\bibitem{Fornal20181}
		Fornal, Bartosz and Grinstein, Benjam\'{\i}n \emph{Dark Matter Interpretation of the Neutron Decay Anomaly}, \emph{Phys. Rev. Lett.} {\bf 120} (2018) 191801.
		
		\bibitem{Fornal_2020}
		Fornal, Bartosz and Grinstein, Benjam\'{\i}n \emph{Neutron’s dark secret}, \emph{Modern Physics Letters A} {\bf 35} (2020) 2030019.
		
		\bibitem{Motta:2018bil}
		Motta, T. F. and Guichon, P. A. M. and Thomas, A. W., \emph{Neutron to Dark Matter Decay in Neutron Stars}, \emph{Int. J. Mod. Phys. A} {\bf 33} (2018) 1844020.
		
		\bibitem{Motta:2018rxp}
		Motta, T. F. and Guichon, P. A. M. and Thomas, A. W., \emph{Implications of Neutron Star Properties for the Existence of Light Dark Matter} \emph{J. Phys. G} {\bf 35} (2018) 05LT01.
		
		\bibitem{refId0}
		Beck, D.H., \emph{Neutron decay, dark matter and neutron stars} \emph{EPJ Web Conf.} {\bf 219} (2019) 005006.
		
		\bibitem{2000NuPhB.564..185M}
		Mielke, E.~W. and Schunck, F.~E.,\emph{Boson stars: alternatives to primordial black} \emph{Nuclear Physics B} {\bf 1} 2000 185-203.
		
		%\bibitem{Abbott_2017}
		%Abbott, B. P. and Abbott, R. and Abbott, T. D. and Acernese, F. and Ackley, K. and Adams, C. and Adams, T. and Addesso, P. and Adhikari, R. X. and Adya, V. B. and et al., %\emph{GW170817: Observation of Gravitational Waves from a Binary Neutron Star Inspiral}, \emph{Physical Review Letters} {\bf 119} (2017) 16.
		
		%\bibitem{Abbott_2019}
		%Abbott, B. P. and Abbott, R. and Abbott, T. D. and Abraham, S. and Acernese, F. and Ackley, K. and Adams, C. and Adhikari, R. X. and Adya, V. B. and Affeldt, C. and et al., %\emph{GWTC-1: A Gravitational-Wave Transient Catalog of Compact Binary Mergers Observed by LIGO and Virgo during the First and Second Observing Runs}, \emph{Physical Review D} {\bf %9} (2019) 3.
		
	\end{thebibliography}
	%\bibliographystyle{ieeetr}

	%\bibitem{a}
	%Author, \emph{Title}, \emph{J. Abbrev.} {\bf vol} (year) pg.
	
	%%\bibitem{b}
	%Author, \emph{Title},
	%arxiv:1234.5678.
	
	%\bibitem{c}
	%Author, \emph{Title},
	%Publisher (year).

	%\end{thebibliography}
	
	% Please avoid comments such as "For a review'', "For some examples",
	% "and references therein" or move them in the text. In general,
	% please leave only references in the bibliography and move all
	% accessory text in footnotes.
	
	% Also, please have only one work for each \bibitem.
	
\end{document}